\documentclass[conference]{IEEEtran}
\IEEEoverridecommandlockouts
\usepackage{cite}
\usepackage{amsmath,amssymb,amsfonts}
\usepackage{algorithmic}
\usepackage{graphicx}
\usepackage{textcomp}
\usepackage{xcolor}
\usepackage{mathtools}
\usepackage{commath}
\usepackage{hyperref}
\hypersetup{colorlinks,allcolors=black}
\usepackage[compact]{titlesec}
\titlespacing{\section}{0pt}{2ex}{1ex}
\titlespacing{\subsection}{0pt}{1ex}{0ex}
\titlespacing{\subsubsection}{0pt}{0.5ex}{0ex}
\usepackage{amsmath}

\def\BibTeX{{\rm B\kern-.05em{\sc i\kern-.025em b}\kern-.08em
    T\kern-.1667em\lower.7ex\hbox{E}\kern-.125emX}}
    
\begin{document}
 
\title{Deep Learning for Partial MIMO CSI Feedback 
by Exploiting Channel Temporal Correlation}

\author{Yu-Chien Lin, Ta-Sung Lee, and Zhi Ding
\thanks{Y.-C Lin is with the Department of Electrical and Computer Engineering,
University of California at Davis, Davis, CA, USA, and
was affiliated with National Yang Ming Chiao Tung University, Taiwan (e-mail: ycmlin@ucdavis.edu).

T.-S Lee is with the Institute of Communications Engineering, National Yang Ming Chiao Tung University, Taiwan (e-mail: tslee@mail.nctu.edu.tw).

Z. Ding is with the Department of Electrical and Computer Engineering,
University of California, Davis, CA, USA (e-mail: zding@ucdavis.edu).}

\thanks{This work is based on materials supported by the National Science Foundation under Grants 2029027 and 2002937 (Ding and Lin) and by the Center for Open Intelligent Connectivity under the Featured Areas Research Center Program within the framework of the Higher Education Sprout Project by the Ministry of Education (MOE) of Taiwan, and partially supported by the Ministry of Science and Technology (MOST) of Taiwan under grant MOST 110-2634-F-009-028, MOST 110-2224-E-A49-001, 110-2622-E-A49-004 and 110-2221-E-A49-025-MY2 (Lee and Lin).}}

\maketitle

\begin{abstract}

Accurate estimation of DL CSI is required to 
achieve high spectrum and energy efficiency in
massive MIMO systems.
Previous works have developed learning-based CSI feedback framework within
FDD systems for efficient CSI encoding and recovery with demonstrated benefits. However, 
downlink pilots for
CSI estimation by receiving terminals
may occupy excessively large number
of resource elements for massive number
of antennas and compromise spectrum efficiency. 
To overcome this problem, we propose a new 
learning-based feedback architecture for efficient encoding of partial CSI feedback of interleaved non-overlapped antenna subarrays by exploiting CSI temporal correlation. For ease of encoding, 
we further design an IFFT 
approach to decouple partial CSI
of antenna subarrays and to preserve partial
CSI sparsity. Our results show 
superior performance in indoor/outdoor scenarios 
by the proposed model for CSI recovery 
at significantly
reduced computation power and storage needs. 
\end{abstract}

\begin{IEEEkeywords}
CSI feedback, temporal correlation, pilot placement, massive MIMO, deep learning
\end{IEEEkeywords}

\section{Introduction}

Much success in gaining high spectrum and energy efficiency has been demonstrated by massive multiple-input multiple-output (MIMO) transceiver systems in
5G and future generations of 
wireless systems. 
Benefits of massive MIMO require 
accurate downlink (DL) channel state information (CSI) at the gNB (i.e., gNodeB). However, in frequency-division duplexing (FDD) systems, 
gNB often relies on user equipment (UE) feedback
to estimate DL CSI. Since the feedback overhead grows proportionally with the increasing array size at the gNB,  CSI feedback reduction is vital to the widespread deployment of FDD MIMO systems.

To improve feedback efficiency, several notable works in \cite{CsiNet, CsiNet+, CsiNetVar3} 
proposed a deep autoencoder framework by deploying
encoder and decoder at UEs and base stations, respectively,
for CSI compression and recovery. They have  
demonstrated significant performance improvement over
traditional compressive sensing approaches. 
Other recent works have revealed the importance of exploiting
correlated channel information from
UL CSI \cite{DualNet, CQNET, DualNet-MP}, past CSI \cite{MarkovNet}, and CSI adjacent UEs \cite{CoCsiNet} to assist DL CSI recovery at base stations. 
For example, important physical insights regarding limited changes in propagation 
environment even under vehicular speed
underscore the strong temporal correlation among 
CSI in delay-angle (DA) domain.
Since
side information from correlated CSI lowers the conditional entropy of the 
DL CSI for recovery,
effective utilization of historical CSI
can reduce encoded feedback payload
required from UEs \cite{MarkovNet}. 

Importantly, the estimation
accuracy of DL CSI at UEs 
depends on several factors such as 
channel fading properties and reference signal (RS)  placement. 
However, the required resource overhead of RS
(i.e. pilot) allocation for CSI estimation 
grows proportionally with the antenna array
size. Too much resource allocated to CSI-RS 
would degrade spectrum efficiency 
Although sparser RS placement can mitigate the problem, it also can lead to larger CSI
distortion due to higher
interpolation error, especially in scenarios of
large multipath delay spread. Yet, to the best of our knowledge, the deep learning frameworks
that take into account RS placement optimization 
to address the tradeoff of CSI accuracy versus spectrum
efficiency has
not been previously investigated for massive MIMO
downlink.  

In this work, we consider a practical RS placement constraint \cite{3GPP} and develop a deep learning (DL) based partial CSI feedback framework which can conserve RS resource overhead while reducing the interpolation loss for high-quality DL CSI 
recovery. Addressing the overall performance of DL CSI recovery, our contributions are as follows:
\begin{itemize}
    \item The work develops an efficient deep learning feedback framework of partial CSI
   by taking RS placement into practical consideration
    
\item We exploit CSI temporal correlation in deep learning network design to achieve improved
CSI recovery performance in terms of accuracy, computation complexity, and memory use.
    
\item We incorporate into the proposed framework
an inverse discrete Fourier transform (IDFT) reformulation for sparser partial CSI representation
and improve the efficiency of
subsequent partial CSI compression.
    
\end{itemize}

\section{System Model}
\subsection{CSI Feedback Framework in FDD system}
Without loss of generality, we
consider a single-cell MIMO FDD link in which
a gNB with $N_b$ antennas communicates with a single
antenna UE. 
The OFDM signal spans $N_f$ DL subcarriers.
The DL received signal of the $k$th subcarrier is 
\begin{equation}
\setlength{\abovedisplayskip}{4pt}
\setlength{\belowdisplayskip}{4pt}
y^{(k)}= \mathbf{h}^{(k)H} \mathbf{w}_\text{T}^{(k)}x^{(k)} + n^{(k)}\label{DL signal},
\end{equation}
where $(\cdot)^H$ denotes  conjugate transpose. 
Here for the $k-$th subcarrier, $\mathbf{h}^{(k)}\in \mathbb{C}^{N_b \times 1}$ denotes the
CSI vector and $\mathbf{w}_\text{T}^{(k)}\in \mathbb{C}^{N_b \times 1}$ denotes the corresponding precoding vector\footnote{gNB calculates precoding vectors at subcarriers with DL CSI matrix.} whereas
$x^{(k)}\in \mathbb{C}$ and $n^{(k)}\in \mathbb{C}$ denote the DL source signal and additive
noise, respectively. 

We rewrite the DL channel vectors as a frequency-spatial CSI (FS-CSI) matrix ${\mathbf{H}}^\text{FS} = [\mathbf{h}^{(1)},...,\mathbf{h}^{(N_f)}]^H \in \mathbb{C}^{N_f \times N_b}$.
Typically in FDD systems, DL CSI  ${\mathbf{H}}^\text{FS}$ is estimated by UE and feedback to gNB. However, the
number ($N_f\times N_b$) of unknowns in  ${\mathbf{H}}^\text{FS}$ occupies
substantial feedback spectrum in large or massive MIMO systems. To exploit CSI sparsity to reduce feedback overhead,
we apply IDFT
$\mathbf{F}_{D}\in \mathbb{C}^{N_f \times N_f}$ and DFT $\mathbf{F}_{A}\in \mathbb{C}^{N_b \times N_b}$ on
$\mathbf{H}^{\text{FS}}$ to generate 
DA domain CSI matrix
\begin{equation}
\setlength{\abovedisplayskip}{4pt}
\setlength{\belowdisplayskip}{4pt}
\mathbf{H}^{\text{DA}} =\mathbf{H}^{\text{DS}}\mathbf{F}_{A}= \mathbf{F}_{D}\mathbf{H}^{\text{FS}}\mathbf{F}_{A},\label{DL CSI matrix in angle-delay domain}
\end{equation}
which demonstrates sparsity. Owing to limited multipath
delay spread and limited number of scatters,
most elements in $\mathbf{H}^{\text{DA}}$
are found to be near insignificant, 
except for the first $Q_f$ and the last $Q_l$ rows.
Therefore, we shorten the CSI matrix $\mathbf{H}^{\text{DA}}$ in DA domain
to $Q_t = Q_f + Q_l$ rows that contain 
sizable non-zero values and 
utilize $\mathbf{H}$ denote the truncated matrices of
${\mathbf{H}}^\text{DA}$. 

Subsequently, to further reduce the feedback overhead, the DL CSI matrix $\mathbf{H}$ is encoded 
at the UE and recovered by the gNB. The recovered DL CSI matrix can be expressed as
\begin{equation}
\setlength{\abovedisplayskip}{3pt}
\setlength{\belowdisplayskip}{3pt}
 \widehat{\mathbf{H}} = f_{\text{de}}(f_{\text{en}}(\mathbf{H})),\label{Recovered DL CSI matrix}
\end{equation}
where $f_{\text{en}}(\cdot)$ and $f_{\text{de}}(\cdot)$ denote encoding/decoding operations.

\subsection{RS placement issue and interpolation loss}
In FDD wireless system, we assume that UEs estimate CSIs based on RSs, termed as pilot CSIs for simplicity, every $T$ time slots which are denoted as \textit{super slots}. Then, UEs interpolate the corresponding DL CSIs of payloads, termed as payload CSIs, according to pilot CSIs for the subsequent data decoding and CSI feedback. In scenarios with CSI changing smoothly along with frequency domain, the interpolation distortion usually can be neglected. However, typically, the number of RS resources need to be proportional to the number of antennas and thus cause fewer resources for data transmission in large scale MIMO system. To avoid this, a simple way is to reduce the RS placement density for each antenna. Yet, an overly sparse RS placement would cause a frequency-selective fading and hence induce a non-negligible interpolation loss. 

\begin{figure}
    \setlength{\abovecaptionskip}{0.cm}
    \setlength{\belowcaptionskip}{-1cm}
\centering
\resizebox{3.4in}{!}{
\includegraphics*{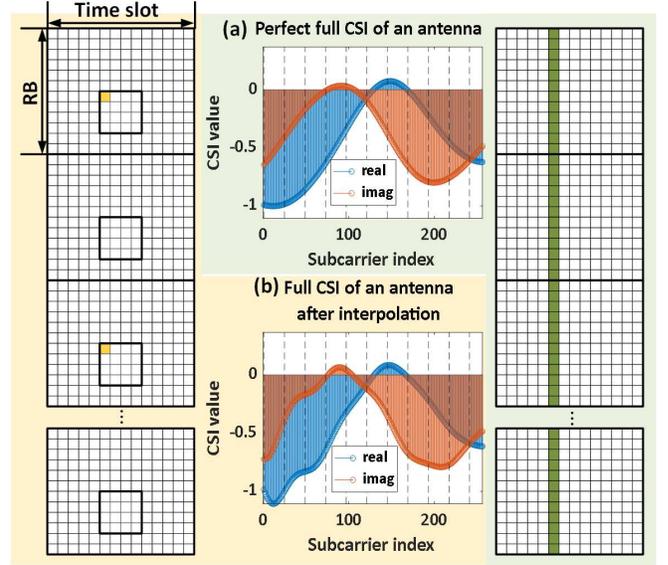}}
\caption{Comparison of (a) perfect CSI and (b) CSI after interpolation according to pilot CSI. (In this example, The simulated channel follows the 3GPP TS38.901 UMa line-of-sight (LOS) channel model. The finest grids are resource elements. Following the specification \cite{3GPP}, the bold black frame covers the region within a resource block (RB) which is assigned to place RSs. Assume that the array size is 32, RSs for each antenna can only be placed every two RBs, causing non-negligible interpolation distortion.)\label{interpolation examples}}
\end{figure}

Assuming that RSs for each antenna are placed at $N_p$ ($N_p \ll N_f$) different subcarriers and the pilot CSIs can be perfectly estimated, we define the $(k,m)$-th element of the raw CSI matrix is given by
\begin{equation}
    (\mathbf{H}^\text{SF, raw})_{\{k,m\}} = \left\{
    \begin{array}{lr}
    (\mathbf{H}^\text{SF})_{\{k,m\}}, &k \in \Omega_m\\
    0, &\text{else}.
    \end{array}
    \right.
\end{equation}
where $(\mathbf{A})_{\{k,m\}}$ denotes the $(k,m)$-th element of a matrix and $\Omega_m$ denotes the subcarrier index set of the $m$-th antenna for RS placement.
The interpolation loss can be expressed as follows:
\begin{equation}
\text{Loss}_\text{itp} = \norm{{\mathbf{H}^\text{DA}_\text{itp}} - {\mathbf{H}^\text{DA}}}^2_\text{F} = \norm{{\mathbf{F}_{D}\widehat{\mathbf{H}}_\text{itp}}\mathbf{F}_{A} - {\mathbf{H}^\text{DA}}}^2_\text{F},
\end{equation}
where $\widehat{\mathbf{H}}_\text{itp} = \text{ITP}(\mathbf{H}^\text{SF, raw})$ is the full CSI after interpolation and $\text{ITP}(\cdot)$ denotes the interpolation operator. As illustrated in Fig. \ref{interpolation examples}, even if we could obtain accurate pilot CSIs, the interpolation loss would be non-negligible when RS placement density is not high enough in some scenarios.

\section{Partial CSI feedback Framework with Spatio-Temporal Division Network (STNet)}
High temporal correlation exists in consecutive time slots and is proved to be used to effectively enhance the CSI recovery performance in previous work \cite{MarkovNet}. However, the RS placement issue and the subsequent interpolation loss are still not tackled. To deal with the interpolation loss, instead of reducing the RS placement density, we propose a learning-based partial CSI feedback framework to estimate and feedback partial CSIs so that the required RS placement density for each antenna could be satisfied and thus a nearly-zero interpolation loss could be achieved.

\begin{figure}
    \setlength{\abovecaptionskip}{0.cm}
    \setlength{\belowcaptionskip}{-1cm}
\centering
\resizebox{3.4in}{!}{
\includegraphics*{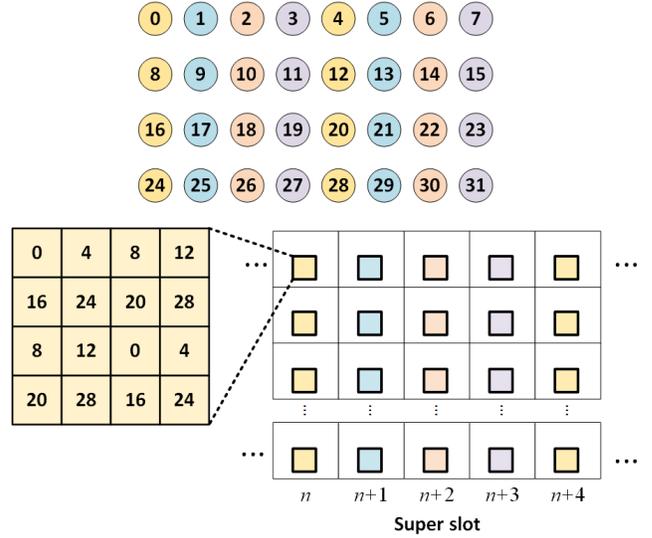}}
\caption{Example of interleaved down-sampled subarray and the corresponding RS placement. (In this example, we consider a MIMO FDD link with the division factor $\kappa = 4$ where the gNB with a uniform planar array (UPA) with $32$ elements attempt to place RSs. To do so, there are two RSs per RB, which can significantly reduce the interpolation loss while being risky in losing some degrees of channel temporal correlation.)\label{array examples}}
\end{figure}

Fig. \ref{basic architecture} shows the general architecture of the proposed framework. Firstly, we separate a whole array into $\kappa$, called \textit{division factor}, non-overlapped and interleaved subarrays as shown in Fig. \ref{array examples} and let UEs to estimate pilot CSIs of the subarrays and do interpolation for payload CSI acquisition. We term the pilot and payload CSI as \textit{partial CSI} collectively. Secondly, an IDFT reformulation pre-processing is applied to partial CSIs for representations with higher sparsity and the combination at the gNB. Then, UEs encode them in every super slots and feedback to the gNB for partial CSI recovery. Subsequently, with the aids of high temporal correlation, a combining network is designed to merge the $\kappa$ partial CSIs received in the most recent $\kappa$ super slots for full CSI recovery. In this section, the above are described in details as follows: 
\begin{figure}
    \setlength{\abovecaptionskip}{0.cm}
    \setlength{\belowcaptionskip}{-1cm}
\centering
\resizebox{3.4in}{!}{
\includegraphics*{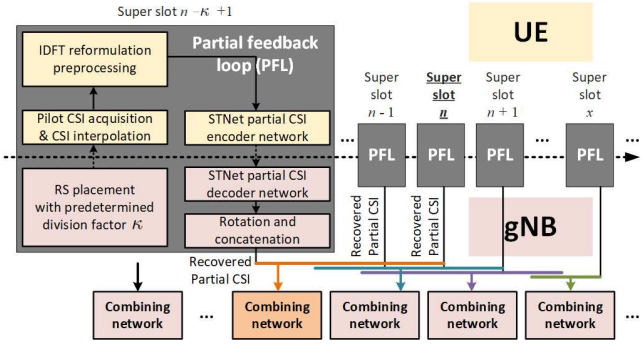}}
\caption{The general architecture of the proposed partial CSI feedback framework. (To estimate the full DL CSI in the $n$-th super slot, the framework requires the current and the previous $\kappa - 1$ partial CSIs. Subsequently, the framework acts like as a sliding window to estimate CSIs of the following super slots. Namely, the required computations related to the previous $\kappa - 1$ feedbacks can be exempted.)\label{basic architecture}}
\end{figure}

\subsection{IDFT reformulation pre-processing}

To maintain the efficiency to encode CSIs, most of the state-of-the-art works \cite{CsiNet, DualNet,CoCsiNet} opt to compress DA domain CSIs at UE side since they are with sparser representations. By reformulating the full CSI, we can decompose the DA domain full CSI as the summation of the $\kappa$ independent partial DA domain CSIs $\mathbf{H}^\text{DA}_\text{0,zp}$,...,$\mathbf{H}^\text{DA}_{\kappa-1\text{,zp}}$ as follows:
\begin{equation}
    \mathbf{H}^\text{DA}_\text{itp}=\mathbf{F}_{D}\widehat{\mathbf{H}}_\text{itp}\mathbf{F}_\text{A}=\mathbf{H}_\text{itp}^\text{DS}\mathbf{F}_\text{A}=\sum_{k=0}^{\kappa-1}{\mathbf{H}^\text{DS}_{k\text{,zp}}\mathbf{F}_\text{A}}
    =\sum_{k=0}^{\kappa-1}{\mathbf{H}^\text{DA}_{k\text{,zp}}},
\end{equation}
\begin{equation}
    (\mathbf{H}^\text{DS}_{k\text{,zp}})_{\{:,m\}} = \left\{
    \begin{array}{lr}
    (\mathbf{H}^\text{DS}_\text{itp})_{\{:,m\}}, &m = k, k+\kappa,...\\
    \mathbf{0}, &\text{else}
    \end{array},
    \right.
\end{equation}
where $\mathbf{H}^\text{DS}$ denotes the CSI in delay-spatial (DS) domain and we can find that the partial DA domain CSIs are periodic with a constant phase difference and a period of $N_b/\kappa$ elements along with the angle axis\footnote{The details of the derivation is described in Appendix.} as shown as follows:

\begin{equation}
    (\mathbf{H}^\text{DA}_{k\text{,zp}})_{\{:,m\}} = (\mathbf{H}^\text{DA}_{k\text{,zp}})_{\{:,m+N_b/\kappa\}}e^{-j2\pi/\kappa}.\label{eq:periodic}
\end{equation}
$(\mathbf{A})_{\{:,m\}}$ denotes the $m$-th column of a matrix $\mathbf{A}$. We denote the truncated matrices of partial CSIs $\mathbf{H}^\text{DA}_\text{0,zp}$,...,$\mathbf{H}^\text{DA}_{\kappa-1\text{,zp}}$ as $\mathbf{H}_\text{0,zp}$,...,$\mathbf{H}_{\kappa-1\text{,zp}}$ in this paper for brevity. In Eq. \ref{eq:periodic}, given the characteristic of the spatially down-sampled subarrays, we can find that only the first $N_b/\kappa$ columns of the truncated matrices $\mathbf{H}_\text{0,zp}$,...,$\mathbf{H}_{\kappa-1\text{,zp}}$ are aperiodic and the remaining are simply their copies with constant phase rotations. Hence, UEs only need to encode the non-periodic portion of the matrices, $\mathbf{H}_\text{0,p}$,...,$\mathbf{H}_{\kappa-1\text{,p}}$, with a size of $Q_t \times N_b/\kappa$, whose size is reduced by a factor of $\kappa$ as compared to the full CSI.  

\subsection{STNet}

We now present a new DLN called STNet. For recovering the CSI of the $n$-th super slot, gNB decodes $\kappa$ codewords received in the current and the previous $\kappa-1$ super slots as $\kappa$ estimated partial CSIs, and combine them as follows:

\begin{equation}
    \widehat{\mathbf{H}}_n = f_\text{C}(\widehat{\mathbf{H}}_{n,\text{ini}}),\label{final_estimate}
\end{equation}
\begin{equation}
    \begin{aligned}
        \widehat{\mathbf{H}}_{n,\text{ini}} = \sum_{k=n-\kappa+1}^{n}\underbrace{\left[
        \begin{matrix}
            \widehat{\mathbf{H}}^T_{k\text{,p}}\\
            \widehat{\mathbf{H}}^T_{k\text{,p}}e^{j2\pi\text{mod}(k,\kappa)/\kappa}\\
            \vdots\\
            \widehat{\mathbf{H}}^T_{k\text{,p}}e^{j2\pi(\kappa-1)\text{mod}(k,\kappa)/\kappa}
        \end{matrix}
        \right]^T}_{\kappa\text{ copies}}
        \label{IDFT combination}
    \end{aligned}
\end{equation}
\begin{equation}
    \widehat{\mathbf{H}}_{k\text{,p}} = f_\text{de}(f_\text{en}(\mathbf{H}_{k\text{,p}}))
\end{equation}
where $f_\text{en}(\cdot)$, $f_\text{de}(\cdot)$ and $f_\text{C}(\cdot)$ denote encoding, decoding, and combining networks, respectively. $\mathbf{H}_{k,p}$ is the non-periodic portion of the partial DL CSI transmitted by the UE in the $k$-th super slot. $\text{mod}(n,k)$ denotes the remainder of a division of $n$ by $k$. Before forwarding to the combining network, due to the periodic features of the partial CSIs, by following Eq. \ref{IDFT combination}, rotation and concatenation operations are applied to the recovered partial CSIs $\widehat{\mathbf{H}}_{k\text{,p}}\text{,} k = m - \kappa +1,...,n$, for initial estimation of the full CSI. Subsequently, it is forwarded to the combining network for the full CSI refinement and estimation.

As for the encoder network, similar to \cite{CQNET}, we forward the real and imaginary parts of CSI to the encoder network, including four $7 \times 7$ circular convolutional layers with 16, 8, 4, and 2 channels and activation functions. Given the circular characteristic of CSI matrices, we introduce circular convolutional layers to replace the traditional linear ones. Subsequently, a fully connected (FC) layer with $\left\lceil{Q_t(N_b/\kappa)/\text{CR}}\right\rceil$ elements is included for dimension reduction after reshaping. $\text{CR}$ denotes the compression ratio. The output of the FC layer is then fed into the quantization module, called the sum-of-sigmoid (SSQ) \cite{CQNET} to generate codewords for feedback.

As for the decoder networks, similar to the encoder, we forward the received codewords to a FC layer with $\left\lceil{Q_t(N_b/\kappa)/\text{CR}}\right\rceil$ elements. Then, four $7 \times 7$ circular convolutional layers with 16, 8, 4, and 2 channels and activation functions are included for partial CSI recovering. Then, the gNB forwards the current recovered partial CSI and the previous estimated $\kappa-1$ partial CSIs to the combining network for final estimation. The combining network adopts the same design of the circular convolutional layers and activation functions as the decoder.

The STNet is optimized by updating the network parameters $\Theta_\text{en}$, $\Theta_\text{de}$ and $\Theta_\text{C}$
\begin{equation}
   \mathop{\arg\min}_{\Theta_\text{en}, \Theta_\text{de}, \Theta_\text{C}}
   \left\{\norm{\widehat{\mathbf{H}}_{n}-\mathbf{H}_{n}}^2_\text{F}
   +
   \sum_{k=n-\kappa+1}^{n}\norm{\widehat{\mathbf{H}}_{k\text{,p}}-\mathbf{H}_{k\text{,p}}}^2_\text{F}/\kappa
   \right\} \\ \label{loss_magnitude}
\end{equation}
where $\mathbf{H}_n$ is the true DL CSI of the $n$-th super slot.

Owing to the framework design, the required storage and computational complexity can be reduced significantly. In details, given the similar features of partial CSIs, only one encoder network needs to be trained and deployed at UEs. Plus, owing to the small input size for partial CSI compression, with the same compression ratio (CR)\footnote{$\text{CR} = \frac{\text{Number of real numbers in CSI to be compressed}}{\text{Number of real numbers in codewords}}$}, the model parameters are reduced with a factor of $\kappa$ as compared to the vanilla CsiNet Pro \cite{MarkovNet}. 
In addition, for recovery of the $n$-th full CSI, although STNet collects the partial CSIs in the most recent $\kappa$ super slots, with the locally available previous $\kappa-1$ partial CSIs, only an additional process to encode and decode the partial CSI in the current super slot is required. 
Thus, the required UE storage and computational complexity can be significantly reduced\footnote{Storage utilization is especially vital for UEs, as compared to gNBs}.
Note that, the selection of the division factor $\kappa$ would significantly affect the recovery performance\footnote{
An excessive $\kappa$ may induce recovery performance degradation since the recovered past partial CSIs could be uncorrelated to the current CSI.}, which will be demonstrated in Section IV. 

\section{Experimental Evaluations}
{
\subsection{Experiment Setup}
In our experiments, we let the UL and DL bandwidths be 20 MHz and the subcarrier number be $N_f$ = 1024. We assume that the FDD system would place CSI-RS every $T = 20$ time slots (i.e., $10$ ms).
We consider both indoor and outdoor cases. 
We place the gNB with a height of 20 m at the center of 
a circular cell coverage
with a radius of 20 m for indoor and 200 m for outdoor. 
We consider a gNB with a $N_b = 32$-element uniform linear array (ULA) communicates with UE with a single antenna. A half-wavelength inter-antenna spacing is considered. 
For each trained model, the number of epochs and batch size were set to 1,000 and 200, respectively. We generate several indoor and outdoor datasets, each containing 100,000 random channels. 
60,000 and 20,000 random channels are for training and validation. 
The remaining 20,000 random channels are test data for performance evaluation.

For indoor data generation, we used the COST 2100 \cite{COST2100} simulator and select the scenario \textit{IndoorHall at 5GHz} to generate indoor channels at 5.1-GHz UL and 5.3-GHz DL with LOS paths. The antenna and band types are set as \textit{MIMO VLA omni} and \textit{wideband}, respectively. As for the outdoor dataset, we utilized QuaDRiGa simulator \cite{QuaDriGa} with the scenario features of \textit{3GPP 38.901 UMa}.
We considered the UMa LOS and NLOS scenarios at 300 and 330 MHz of UL and DL bands, respectively, with different UE velocities (considering 1, 2 and 4 m/s). The antenna type is set to \textit{omni}. 

The performance metric is the normalized MSE
\begin{equation}
\text{NMSE} =  \frac{1}{D}\sum^{D}_{d=1}\norm{\widehat{\mathbf{H}}_{d}^\text{SF} - {\mathbf{H}}_{d}^\text{SF}}^2_\text{F} /\norm{{\mathbf{H}}_{d}^\text{SF}}^2_\text{F},\label{NMSE1}
\end{equation}
where the number $D$ and subscript $d$ denote the total number and index of channel realizations, respectively. Instead of evaluating the estimated DL CSI matrix $\widehat{\mathbf{H}}$, we evaluate the estimated SF-CSI matrix $\widehat{\mathbf{H}}^\text{SF}$ that can be obtained by reversing the Fourier processing and padding zero matrix.
Note that this NMSE includes both the
errors caused by truncation at the encoder
and the overall recovery error. Thus, it
is practically more meaningful. 

In the following section, we evaluate the performance 
of CSI recovery by adopting the proposed partial CSI feedback framework. To show the efficacy of the framework, we compare the STNet with CsiNet Pro used in \cite{MarkovNet} which shares the same core layer design of the encoder/decoder networks. Namely, the CsiNet Pro can be regarded as the STNet without multiple super-slot branches and the combing network. As shown in Fig. \ref{layer architecture comparison}, the only difference between STNet and CsiNet Pro is the IDFT reformulation pre-processing and the additional combining network.

As mentioned in Section III, STNet can obtain the current full CSI by obtaining the current partial CSI. Thus, in each super slot, as compared to CsiNet Pro, the codeword size to be transmitted at the same CR can be reduced by a factor of $\kappa$. Herein we define a new term called effective compression ratio ($\text{CR}_\text{eff} = \kappa\text{CR}$).

\begin{figure}
    \setlength{\abovecaptionskip}{0.cm}
    \setlength{\belowcaptionskip}{-1cm}
\centering
\resizebox{3.4in}{!}{
\includegraphics*{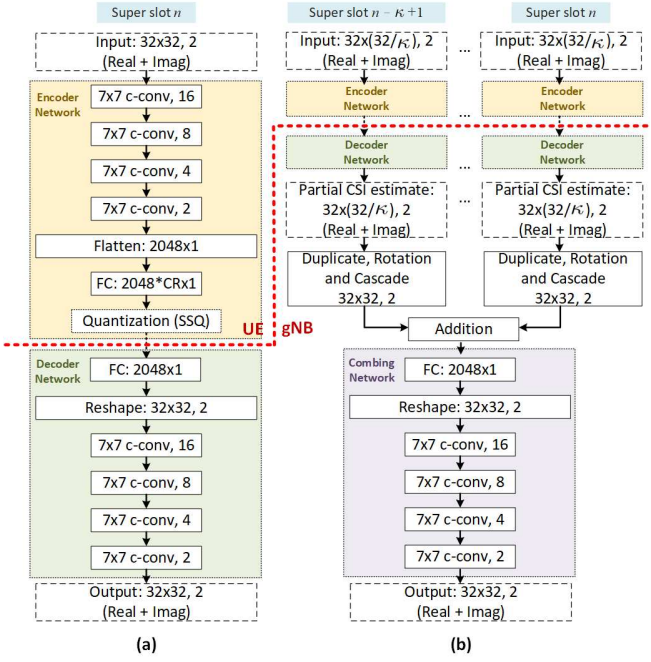}}
\caption{Architecture of (a) CsiNet Pro and (b) STNet.\label{layer architecture comparison}}
\end{figure}

\subsection{Different Division Factor $\kappa$}

To demonstrate the superiority of the proposed framework, we first applied different division factors $\kappa$ in both indoor and outdoor scenarios with low UE mobility ($v = 1$m/s) for different effective compression ratios $\text{CR}_\text{eff}$. Fig. \ref{sim1} (a), (b) and (c) show the NMSE performance of the proposed framework STNet and vanilla CsiNet Pro under indoor and outdoor scenarios, respectively, at different compression ratios. By adopting the proposed framework, with the aids of the high CSI temporal correlation, STNet performs much better than CsiNet Pro and has the best performance when using a larger division factor $\kappa$. 
However, for higher recovery accuracy, we still cannot just choose a large division factor $\kappa$ in all scenarios, especially in outdoor scenarios since STNet highly depends on the CSI temporal correlation. The reason will be shown in the following section.


\begin{figure}
\centering
\resizebox{3.4in}{!}{
\includegraphics*{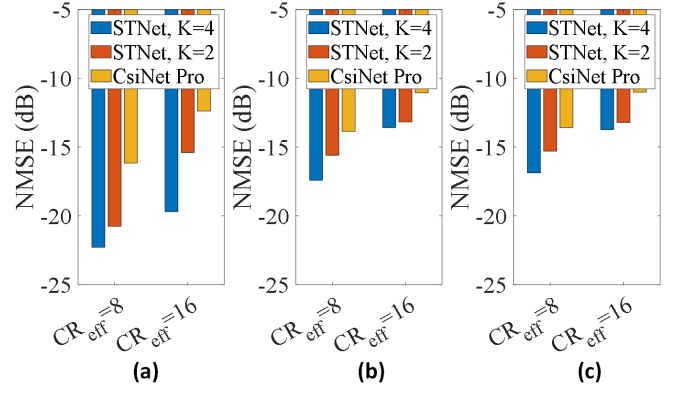}}
\caption{NMSE performance for different division factor $\kappa$ in (a) indoor, (b) outdoor UMa LoS and (c) outdoor UMa NLoS scenarios.\label{sim1}}
\end{figure}

\subsection{Different UE mobility}
To comprehensively assess the robustness of the proposed framework, we tested the same alternatives in outdoor scenarios with different UE mobility (= 1, 2 , 4 m/s\footnote{These velocities roughly corresponds to the UE velocities of walking, jogging, and biking, respectively.}). Fig. \ref{sim2} shows the NMSE performance of STNet and CsiNet Pro in the outdoor scenarios with different average UE velocity and at different compression ratios. We can find that STNet performance degrades as the increasing UE velocity and even perform worse than CsiNet Pro when division factor $\kappa = 4$. Namely, the higher UE mobility would induce a lower CSI temporal correlation, causing the performance degradation of STNet. According to the numerical simulations, to strike a trade-off, an intermediate division factor $\kappa = 2$ seems to be a plausible solution in outdoor scenarios.

\begin{figure}
\centering
\resizebox{3.4in}{!}{
\includegraphics*{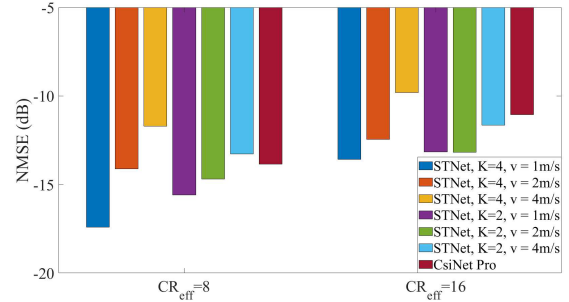}}
\caption{NMSE performance for different UE velocities in the outdoor UMa LoS scenario.\label{sim2}}
\end{figure}
\subsection{Assessment of FLOPs and parameter numbers}


\begin{table}
\centering
\caption{FLOP and parameter numbers of STNet and CsiNet Pro.\label{Complexity_Diff_Encoding}}
\begin{tabular}{|c|c|c|c|c|}
\hline
      & \multicolumn{2}{c|}{FLOPs} & \multicolumn{2}{c|}{Parameters} \\ \hline
Method & $\text{CR}_\text{eff}=8$            & $\text{CR}_\text{eff}=16$         & $\text{CR}_\text{eff}=8$            & $\text{CR}_\text{eff}=16$              \\ \hline
STNet, $K=2$  & 1.1M        & 583K       & 555K           & 292K           \\ \hline
STNet, $K=4$  & 583K        & 321K       & 292K           & 160K           \\ \hline
CsiNet Pro  & 2.13M        & 1.08M       & 1.07M          & 0.54M          \\ \hline
\end{tabular}
\end{table}
Due to the limited storage and computation power of UEs, it is crucial to design a light-weighted and easy-to-implement neural network. Table \ref{Complexity_Diff_Encoding} shows that, besides the superiority in terms of the NMSE performance, the required computation power and storage for model parameters can also significantly reduced. To explain, with smaller input and output sizes of STNet's encoder and decoder networks, the required FLOPs and parameters are significantly lower than CsiNet Pro. In addition, since STNet can act in a sliding window fashion, the encoder and decoder networks are conducted only once instead of $\kappa$ times. Thus, even considering the implementation of the combing network, the STNet's FLOPs and parameters are still obviously less than the ones of CsiNet Pro.}
\section{Conclusions}
This work presents a new deep learning framework 
for large scale CSI estimation 
that utilizes partial CSI feedback compression
and leverages CSI temporal correlation in FDD systems.
To control time-frequency resources needed for CSI-RS in the proposed IDFT reformulation, our
proposed STNet allows UE to transmit
partial CSI feedbacks associated with interleaved non-overlapped subarrays in $\kappa$ super slots.
Subsequently, the gNB estimates the full CSI based on aggregated partial feedbacks by utilizing
CSI temporal correlation. Our test results 
demonstrate significant improvement
of recovery performance at
reduced computation complexity and storage
usage in comparison with the existing
benchmark model of CsiNet Pro.
\section*{Appendix}
\setcounter{equation}{0}
\setcounter{subsection}{0}
\renewcommand{\theequation}{A.\arabic{equation}}
\renewcommand{\thesubsection}{A.\arabic{subsection}}
\subsection*{Proof of Eq. (\ref{eq:periodic}):}
Since the array is divided into non-overlapped and interleaved subarrays, the CSI matrix in DS and DA domains for the $k$-th subarray are expressed respectively as follows:  
\begin{equation}
    (\mathbf{H}^\text{DS}_{k\text{,zp}})_{\{:,m\}} = \left\{
    \begin{array}{lr}
    (\mathbf{H}^\text{DS}_\text{itp})_{\{:,m\}}, &m = k, k+\kappa,...\\
    \mathbf{0}, &\text{else}.
    \end{array}
    \right.
\end{equation}

\begin{equation}
\begin{aligned}
(\mathbf{H}^\text{DA}_{k\text{,zp}})_{\{:,m\}} & =
(\mathbf{H}^\text{DS}_{k\text{,zp}}\mathbf{F}_\text{A})_{\{:,m\}} = \mathbf{H}^\text{DS}_{k\text{,zp}}(\mathbf{F}_\text{A})_{\{:,m\}}\\
&= \sum_{m'=k,k+\kappa,...}(\mathbf{H}^\text{DS}_\text{itp})_{\{:,m'\}}w^{(m'-1)},
\end{aligned}
\end{equation}
By substituting the column index $m$ by $m^*=m+N_b/\kappa$, we can find that the CSI matrix in DA domain is periodic with a constant phase difference and a period of $N_b/\kappa$ elements along with the angle axis as follows:
\begin{equation}
\begin{aligned}
(\mathbf{H}^\text{DA}_{k\text{,zp}})_{\{:,m^*=m+N_b/{\kappa}\}} & =
(\mathbf{H}^\text{DS}_{k\text{,zp}}\mathbf{F}_\text{A})_{\{:,,m^*\}} \\ &= \mathbf{H}^\text{DS}_{k\text{,zp}}(\mathbf{F}_\text{A})_{\{:,,m^*\}}\\
& = \sum_{m=k,k+\kappa,...}(\mathbf{H}^\text{DS}_\text{itp})_{\{:,m\}}w^{(m+N_b/\kappa-1)}\\
& = (\mathbf{H}^\text{DA}_{k\text{,zp}})_{\{:,m\}}e^{-j2\pi/\kappa}
\end{aligned}
\end{equation}
where 

\begin{equation}
    (\mathbf{F}_\text{A})_{\{:,m\}} = [1,w^{(m-1)},...,w^{(N_b-1)(m-1)}]^T,\label{eq:IDFT}
\end{equation}
\begin{equation}
    w = e^{-j2\pi/N_b}.\label{eq:IDFT factor}
\end{equation}
\bibliography{references.bib}
\bibliographystyle{IEEEtran}
\end{document}